# Design of Mechanical Structure and Tracking Control System for 5 DOF Surgical Robot

S.M. Sajadi[*], S.H. Mahdioun[**], A.A. Ghavifekr[***]

* Department of Mechatronic Engineering, School of Engineering-Emerging Technologies, University of Tabriz, Tabriz, Iran
Email: mr.sajadi89@ms.tabrizu.ac.ir
** Department of Mechatronic Engineering, School of Engineering-Emerging Technologies, University of Tabriz, Tabriz, Iran
Email: h.mahdioun90@ms.tabrizu.ac.ir
*** Department of Electrical and Computer Engineering, University of Tabriz, Tabriz, Iran
Email: a_aminzadeh89@ms.tabrizu.ac.ir

**Abstract:**. *In this paper, the tracking control problem for 5 DOF surgical robot which is affected by tremor of surgeons' hand is considered. Mechanical modeling and dynamic analysis of a robotic arm in slave subsystem of a telesurgery system will be discussed and the reasons for selecting the appropriate materials for different parts of robot will be explained. It would be required this robot, which will do the main part of the surgery, be controlled based on the uncertain properties of the tissues of patients body.*
*Improved Lyapunov Based control method with uncertainty observer is applied to improve the accuracy of tracking procedure for a surgical manipulator to track a specified reference signal in the presence of tremor that is modeled as constant bounded disturbance. Based on the disturbance rejection scheme, tracking controllers are constructed which are asymptotically stabilizing in the sense of Lyapunov. The control strategy was implemented using a PC interface. Computer simulation results demonstrate that accurate trajectory tracking can be achieved by using the proposed controllers.*

## 1. Introduction

Traditional surgical procedures include creating a large hole in the body, in order to access the desired tissue and performing the desired operation. This type of surgery is called open surgery method [1]. The next innovative method was MIS which is also called laparoscopic operation [2]. The aim of robot aided surgery is to use an instrument which utilized robotic arms that are installed on patient's body and reflect the movements of surgeon's hands safely and accurately into the patient's body [3]. The fundamental of robotic surgery is based on laparoscopic surgery method [4]. In laparoscopic surgery, camera and surgical instruments are sent into patient's body by creating several small holes and trajectory of this motion is controlled and investigated by surgeon. In this method, because of the small diameter of the holes, in order to pass surgical instruments with 5 to 15 mm diameters interval, convalescence, hospitalization period and surgical costs are dramatically reduced [2]. This provides an appropriate background for using surgical robots with the aim of minimally invasiveness. In the two recent decades, widespread efforts have been done in order to construct surgical robots. Two of the major achievements in this field are "ZEUS" and "DAVINCI" surgical robots [5]. Telesurgery is one the most important applications of surgical robots, in which there are two subsystem, master and slave, which are connected with each other and with patient's body in desired spatial distances [6,7].

These robots not only increase the accuracy and simplicity of surgery, but also avoid unwanted hurts to the patients' body by using haptic structure and taking force feedback [8]. One of the important parts of surgical robots is their grasper which is in contact with the tissues of the patients' body. Making autonomous graspers of surgical robots would play a major role in reducing the errors of the surgical operations and would improve the surgeons' perception of the region of operation.

The robotic arm which is used in slave subsystem should be designed according to the aim of eliminating surgeon's ergonomic problems. The proposed robot is an optimized version of compact laparoscopic surgical robotic system that is constructed in [9]. In this work, ball-screw mechanism is used instead of rack and pinion to improve accuracy of robot's performance which is the most important movement in this robot [10]. One of the most important points in slave structure is to cover the required workspace by robot's End-Effector. The constructed robot uses spherical coordinate which covers the required workspace by surgeon using three degrees of freedom in order to place the End-Effector in the desired position [11]. One degree of freedom is for creating the roll movement in order to place the instrument's End-Effector in an accurate position, and one degree is assigned for laparoscopic instrument gripper in robot's workspace.

The tracking control of robotic manipulators has been extensively studied. Trajectory tracking errors for robotic system are subjected to various disturbances, such as measurement and modeling error and load variance [12]. In order to obtain better tracking performance the developed control algorithms should be capable of reducing these uncertainties effects.

If the parameters of the robot are completely known, feedback linearization or computed torque scheme can be used for control design. When system parameters are unknown and vary in wide ranges, adaptive and robust control algorithms need to be applied [13].

Tremor of surgeons' hands and hysteresis phenomena are two major problems in controlling the end-effectors of surgical robots that lead to considerable errors. To compensate hand tremor, there are two approaches: tremor model based and robust approach. In this paper, the second method will be used to deal with the problem of eliminating the effect of hand tremors of surgeon.

The effect of these tremors in dynamic equations can be represented by a term summed with systems input, which is called disturbance in control literature. In [14] three nonlinear control methods are evaluated to reject constant bounded disturbances from robotic manipulators. Inverse-dynamic controller with integral action is an example of methods to eliminate the additive terms that are used to cancel uncertain constant-bounded disturbances like load torque. In this paper Lyapunov-based controller with uncertainty observer is used which can eliminate time-variant and bounded disturbances, considering tremor of surgeon's hand properties.

The rest of this paper is organized as follows: In section 2 physical and mechanical structure of designed surgical robot is explained. This section describes dynamic model and geometric calculation of grasper. In section 3 robotic tracking problem is formulated into a proper tremor rejection problem and two nonlinear control methods to reject the tremor of surgeons' hand that is considered as constant bounded disturbances are presented. All system uncertainties are lumped into the disturbance term. In order to prove the effectiveness of the proposed controllers, simulation results have been presented in section 4.

## 2. Mechanical Modeling and Dynamics Analysis of the Manipulator

In designing part of robot, first an initial model of the robot is developed and the required materials and tools are selected.

### 2.1 Mechanical Modeling and Construct

Taking the parameters such as reduction in dimensions, weight, cost, installation, maintenance and sterilization of the robotic arm into consideration, the proposed robot is an optimized version of compact laparoscopic surgical robot system that constructed in [9]. To improve the accuracy, ball-screw mechanism is used instead of rack and pinion mechanism in the optimized model. In order to reduce the weight of the manipulator, the main material of robot's chassis is chosen to be aluminum. Cutting of aluminum is done by abrasion Water-Jet cutting method [15]. So that after cutting process, machining process will be done. The material of the used gears is steel 316 which has the advantage of good sterilization [16]. According to needed accuracy and to eliminate backlash the gears are constructed using Wire-Cutting method. Fig.1 shows the designed manipulator.

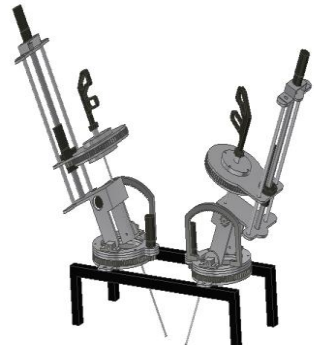

Fig.1. Mechanical Simulation of manipulator

Three sets of sun gear are used for base azimuth rotation unit, supporting rod and insertion and instrument driven unit in order to increase motors' output torques and decrease dimensions and weights. Table 1 illustrates the used set of gears with detail.

Table 1. Detailed part of the robot

| Position | Schematic figure | Gear Ratio |
|---|---|---|
| Azimuth Rotation Unite | 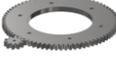 | 7.7:1 |
| Supporting Rod | 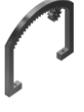 | 10.2:1 |
| Insertion and Instrument Driven Unite | 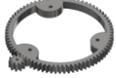 | 7.7:1 |

The rotation of base azimuth rotation unit is done using 6 ball-bearings. The rotation of supporting rod is done using 2 thrust ball-bearing and a ball-bearing is used to ease the rotation of insertion and instrument driver unit. For translation a ball-screw mechanism is used. This mechanism reduces the weight of selected motors for this robot and increases the accuracy of the translation. After completion of design and simulation stage, selection of motors and other components and also selection of construction method have been considered. The slave manipulator has been shown in Fig. 2.

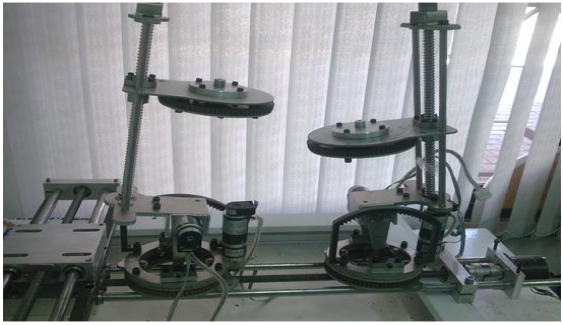

Fig. 2. The slave manipulator constructed

### 2.2 Manipulator Dynamic Model

Dynamic modeling of surgical robots has two general methods: Energy-based methods and the methods based on Newton formulation. Methods like Lagrange and Hamiltonian that are based on energy conservation law are in the first category. But methods like Cosserat rod theory that are based on Newton motion equations are in the second category. To determine position of the robots' end-effector in Cartesian coordinate relative to ground, the Denavit Hartenberg [17] parameters can be used. Using Lagrange method, dynamical model of the manipulator is derived from a simplified model that is shown in Fig.3. The details of following notations have been mentioned in Table 2.

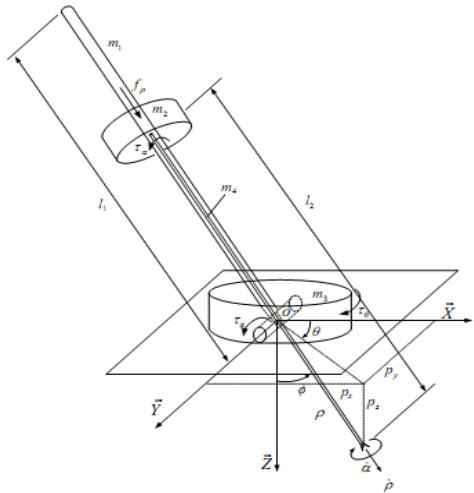

Fig. 3. Simplified Manipulator Model

By considering 4 degrees of freedom, this manipulator has 4 generalized coordinates $q_1, q_2, q_3, q_4$ which corresponded to joint position parameters $\theta, \phi, \alpha, \rho$ respectively. $T$ is kinetic energy; $V$ is potential energy and $Q_i$ is generalized force which corresponded to joints torque and force $\tau_\theta, \tau_\phi, \tau_\alpha, \tau_\rho$. Table 2 shows the parameters obtained from mechanical modeling and parameters from the prototype of the Manipulator.

Table 2. parameters obtained from mechanical model

| Part Name | Picture of Part with Center of Gravity | Specifications |
|---|---|---|
| Supporting Rod | | $M_1$=1.541kg<br>$I_{1a}$=32045.478 kg.mm^2<br>$I_{1t}$=31429.513 kg.mm^2<br>$L_1$=520mm |
| Insertion and Instrument Driven unite | | $M_2$=1.613kg<br>$I_{2a}$=6317.537 Kg.mm^2<br>$I_{2t}$=2401.198 Kg.mm^2 |
| Base Azimuth Rotation Unit | | $M_3$= 0.915 kg<br>$I_{3a}$=4249.517 Kg.mm^2 |
| Surgical Instrument with its Accessories | | $M_{4(only\ instrument)}$= 0.089kg<br>$I_{4a(instrument\ with\ its\ accessories)}$= 2681.116 kg mm^2<br>$I_{4t\ (only\ instrument)}$= 1358.560 kg mm^2<br>$L_2$=300mm |

Kinetic energy of the Manipulator include of rotational and translational kinetic energies.
Rotational kinetic energy of the supporting rod:

$$T_1 = \frac{1}{2}I_{1a}\left(\dot{\theta}\cos\phi\right)^2 + \frac{1}{2}\left[I_{1t} + m_1\left(\frac{l_1}{2}\right)^2\right]\left[\dot{\phi}^2 + \left(-\dot{\theta}\sin\phi\right)^2\right] \quad (1)$$

Rotational and translational kinetic energy of the insertion driver unit:

$$T_2 = \frac{1}{2}I_{2a}\left(\dot{\theta}\cos\varphi\right)^2 + \frac{1}{2}\left[I_{2t} + m_2\left(l_2 - \rho\right)^2\right]\left[\dot{\phi}^2 + \left(-\dot{\theta}\sin\phi\right)^2\right] + \frac{1}{2}\left(m_2\dot{\rho}^2\right) \quad (2)$$

Rotational kinetic energy of the azimuth base unit:

$$T_3 = \frac{1}{2} I_{3a} \dot{\theta}^2 \tag{3}$$

Rotational and translational kinetic energy of the instrument:

$$T_4 = \frac{1}{2} I_{4a} \left( \dot{\theta} \cos\varphi + \dot{\alpha} \right)^2 +$$
$$\frac{1}{2}\left[ I_{4t} + m_4 \left( \frac{l_2}{2} - \rho \right)^2 \right] \left[ \dot{\phi}^2 + (-\dot{\theta}\sin\phi)^2 \right] + \frac{1}{2} m_4 \dot{\rho}^2 \tag{4}$$

The total kinetic energy of the system is:

$$T = T_1 + T_2 + T_3 + T_4 \tag{5}$$

Potential energy of the system is caused by gravity and includes supporting rod potential energy and insertion driven unit potential energy.

$$V = m_1 g \frac{i_1}{2} \cos\phi + m_2 g (l_2 - \rho) \cos\phi +$$
$$m_4 g \left( \frac{i_2}{2} - \rho \right) \cos\phi \tag{6}$$

By solving Lagrange equation for $\theta, \phi, \alpha, \rho$ generalized coordinates, the dynamical equation of the Manipulator is as follows:

$$\tau = M\ddot{q} + c[q, \dot{q}] + G + R \tag{7}$$

where $M(q) \in R^{n \times n}$ is the inertia matrix, $C(q,\dot{q})\dot{q} \in R^n$ is the centripetal and Coriolis matrix, $g(q) \in R^n$ is the gravitational force and u is the exerted joint input. $q \in R^n$ is the joint angle vector and R is Friction matrix.

### 3. Control Law Formulation

Tremor of surgeon or operator's hand can be considered as a disturbance input, whose effect can be eliminated by designing an appropriate controller. In this section it will be tried to find a solution to this problem. The surgeon's hand tremor will be studied as a constant-bounded disturbance. This disturbance will be considered as load torques on input signals. More specifically if $d$ be a constant-bounded vector, then the dynamic equation of motion for the surgical robot can be written as :

$$M\ddot{q} + C\dot{q} + G(q) = u + d \tag{8}$$

Lyapunov-based controller is a type of passivity-based controllers and unlike inverse dynamic method does not rely on linearization or equation decoupling. This nonlinear controller uses a passive mapping to filter trajectory tracking error and produces a control law based on this derived variable.

As a first step, assume the following variable substitution:

$$\begin{cases} \dot{\xi} = \dot{q}_d - F(\circ)\tilde{q} \\ \sigma = \dot{q} - \dot{\xi} \end{cases} \tag{9}$$

$$\rightarrow \sigma = \dot{\tilde{q}} + F(\circ)\tilde{q} \xrightarrow{Laplace} \sigma(s) = (sI + F(s))\tilde{q}(s)$$

$F(.)$ can be assumed as any linear operator, in the condition that mapping $H(s) = (sI + F(s))^{-1}$ be strictly proper and stable. Indeed $H(s)$ is an operator which maps trajectory tracking error to a new variable $\sigma$. Based on this fact, Lyapunov-based controller law with minor correction can be derived as:

$$u := M(q)\ddot{\zeta} + C(q,\dot{q})\xi + G(q) - K_D\sigma - \hat{d} \tag{10}$$

where $\hat{d}$ is an estimate of $d$. If $\hat{d} = d$, then disturbance will be rejected. Now let consider $\hat{d} \neq d$ thus

$$\tilde{d} = d - \hat{d} \tag{11}$$

Here, $\tilde{d}$ is the estimation error. Using (11) in (8) yields

$$M(q)\dot{\sigma} + C(q,\dot{q})\sigma + K_D\sigma = \tilde{d} \tag{12}$$

According to passivity theorem [18], if mapping of $-\sigma \mapsto \tilde{d}$ is passive relative to the some functions of $V_1$ and also $\tilde{d}$ is bounded, then $\tilde{q}$ will be continuous and $\tilde{q}, \dot{\tilde{q}}$ will asymptotically converge to zero. It means that

$$\lim_{t \to \infty} \tilde{q}(t) = \lim_{t \to \infty} \dot{\tilde{q}}(t) = 0 \tag{13}$$

First it is shown that $\tilde{d}$ is bounded and then appropriate estimation law is proposed for $\hat{d}$. Let define V as

$$V := \frac{1}{2}\sigma^T M(q)\sigma + \frac{1}{2}\tilde{d}^T K_I^{-1}\tilde{d} \tag{14}$$

where $K_I$ is a positive definite symmetric matrix. The time derivative of V is given by

$$\dot{V} = \sigma^T M(q)\dot{\sigma} + \frac{1}{2}\sigma^T \dot{M}(q)\sigma + \tilde{d}^T K_I^{-1}\dot{\tilde{d}}$$
$$= \sigma^T (-C(q,\dot{q})\sigma - K_D\sigma + \tilde{d}) + \frac{1}{2}\sigma^T \dot{M}(q)\sigma + \tilde{d}^T K_I^{-1}\dot{\tilde{d}} \tag{15}$$
$$= \frac{1}{2}\sigma^T (\dot{M}(q) - 2C(q,\dot{q}))\sigma - \sigma^T K_D\sigma + \tilde{d}^T (\sigma + K_I^{-1}\dot{\tilde{d}})$$
$$= -\sigma^T K_D\sigma + \tilde{d}^T (\sigma + K_I^{-1}\dot{\tilde{d}})$$

Now suppose that

$$\sigma + K_I^{-1}\dot{\tilde{d}} = 0 \rightarrow \dot{\tilde{d}} = K_I\sigma \tag{16}$$

With substituting (16) in (15)

$$\dot{V} = -\sigma^T K_D\sigma \leq 0 \tag{17}$$

Since $V$ is bounded from below ($V \geq 0$) and decreasing ($\dot{V} \leq 0$), then $\lim_{t \to \infty} V(t)$ is bounded too.

Since $\frac{1}{2}\sigma^T \dot{M}(q)\sigma$, $\frac{1}{2}\tilde{d}^T K_I^{-1}\dot{\tilde{d}}$ are non-negative matrices and $M(q)$, $K_I^{-1}$ are limited matrices, then $\sigma$ and $\tilde{d}$ are bounded. It means that $\sigma, \tilde{d} \in L_\infty$. By using

$$V(t) - V(0) \leq -\lambda_{\min}(K_D) \int_0^t \|\sigma(s)\|^2 ds \tag{18}$$

It is achieved that $\sigma \in L_2$.

Boundary of $\sigma$ leads to bounded $\tilde{q}, \dot{\tilde{q}}$, thus $C(q,\dot{q})$ is bounded and (19) is obtained.

$$\dot{\sigma} \in L_\infty$$
$$\dot{V} = -\sigma^T K_D \sigma \to \ddot{V} = -2\sigma^T K_D \dot{\sigma} \quad (19)$$
$$\sigma, \dot{\sigma} \in L_\infty \to \ddot{V} \in L_\infty$$

By using Barbalat's lemma [18] it can be shown that

$$\begin{cases} \ddot{V} \in L_\infty \\ \lim_{t\to\infty} V(t) < \infty \end{cases} \to \lim_{t\to\infty} \dot{V}(t) = 0 \to \lim_{t\to\infty} \sigma = 0 \quad (20)$$

Update law that was mentioned in (16) leads to

$$-\sigma^T \tilde{d} = -\dot{\hat{d}}^T K_I^{-1} \tilde{d} = \dot{\tilde{d}}^T K_I^{-1} \tilde{d}$$

$$\int_0^t -\sigma^T(s)\tilde{d}(s)ds = \int_0^t \dot{\tilde{d}}^T(s) K_I^{-1}\tilde{d}(s)ds = \frac{1}{2}\tilde{d}^T K_I^{-1} \tilde{d} \Big|_0^t \quad (21)$$

$$= \frac{1}{2}\tilde{d}^T K_I^{-1} \tilde{d} - \frac{1}{2}\tilde{d}^T(0) K_I^{-1} \tilde{d}(0) = V_1(t) - V_1(0)$$

where $V_1 = \frac{1}{2}\tilde{d}^T K_I^{-1} \tilde{d}$. Thus, mapping of $-\sigma \mapsto \tilde{d}$ is passive relative to the some functions of $V_1$ and $\lim_{t\to\infty}\tilde{q}(t) = \lim_{t\to\infty}\dot{\tilde{q}}(t) = 0$.

Control law that was defined in (11) can be rewritten as

$$u = M(q)\ddot{\zeta} + C(q,\dot{q})\xi + G(q) - K_D \sigma - K_I \int_0^t \sigma(s)ds \quad (22)$$

By comparing inverse-dynamics and Lyapunov – based controllers, it can be stated that inverse-dynamic based controller has the capability to be modified, while Lyapunov-based controller cannot be easily modified due to its structural complexities. But from another viewpoint, the Lyapunov-based controller is efficiently robust to uncertainties due to its nonlinear nature, where inverse dynamic based controller is very sensitive to parameter uncertainties.

## 4. Simulation

The control goal is tracking desired trajectory in joint and work space in the presence of surgeon's hand tremor. Tremor modeling leads to terms that are added with the input. Lyapunov-based controller with uncertainty observer is the controller which will be discussed in (10). Control law based on Lyapunov method in workspace can be written as:

$$\begin{cases} u = M(q)\ddot{\xi} + C(q,\dot{q})\dot{\xi} + G(q) + J_a^T(q) K_D J_a(q)(\dot{\xi} - \dot{q}) \\ \dot{\xi}_x = \dot{x}_d + \Lambda(x_d - x) \quad , \quad \dot{\xi} = J_a^\dagger \dot{\xi}_x \end{cases} \quad (23)$$

In above equation, matrices $K_d, \Lambda$ are symmetric positive definite. To design control law, gain matrices are assumed to be $K_D = 4 \times I_2$, $\Lambda = 2 \times I_2$.

The reference trajectory of controller should be described in the workspace. Designed surgical robot has two singularity in k = 0 and 2.3311. The k = 0 singular point is a marginal point in robots achievable workspace, so is not considered to be a critical singular point in practice but the control law can experience major variations in some moments.

To resolve this problem it is possible to simply avoid singular configurations by choosing a singular point free trajectory. To do this, by solving forward and inverse kinematics for robot, following equations can be derived:

$$\begin{cases} x_d = \cos(t)/2 + \sin(\sin(t)/2 + 1)/(\sin(t)/2 + 1) \\ y_d = -(\cos(\sin(t)/2 + 1) - 1)/(\sin(t)/2 + 1) \end{cases}$$

$$\begin{cases} dx_d = (\cos(\sin(t)/2+1) \times \cos(u))/(2 \times (\sin(t)/2+1)) + \ldots \\ \quad -\sin(t)/2 - (\sin(\sin(t)/2+1) \times \cos(t))/(2 \times (\sin(t)/2+1)^2) \\ dy_d = (\sin(\sin(t)/2+1) \times \cos(t))/(2 \times (\sin(t)/2+1)) + \ldots \\ \quad +(\cos(t) \times (\cos(\sin(t)/2+1)-1))/(2 \times (\sin(t)/2+1)^2) \end{cases}$$

(24)

This trajectory does not enter singular configuration in any moment. Now performance of Lyapunov-based controller in tracking a singular point free trajectory will be discussed. For this reference trajectory, Figure.4 represents trajectory tracking of Lyapunov-based controller in workspace and the corresponding control signals.

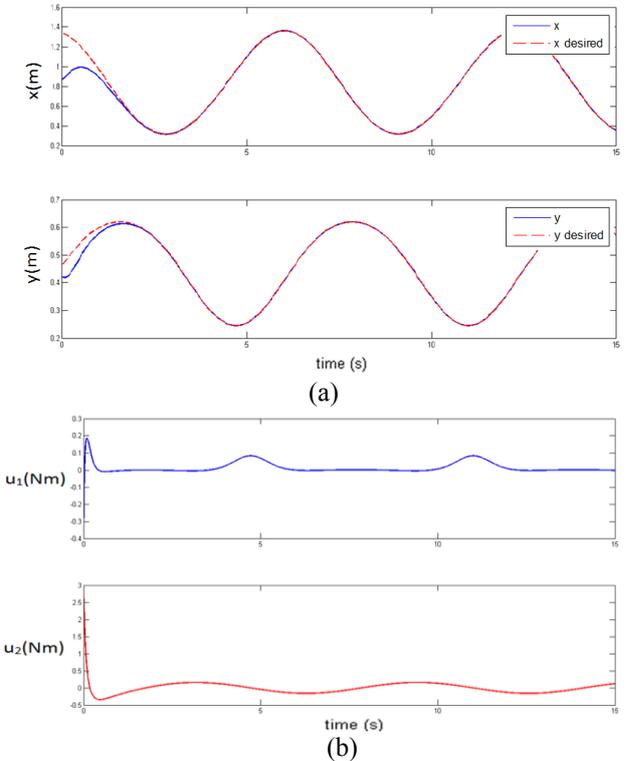

Fig. 4. a) Tracking performance of Lyapunov-based controller in workspace, b) Control signals of Lyapunov-based controller

This controller is robust to the magnitude of uncertainty and it is not necessary to modify its gains. In order to compare this controller with inverse-dynamic controller with an integral action, an uncertainty with a magnitude 10 times larger $\Delta \tau = [10 \ 10]^T$ is imposed to this controller. Utilized control gains are $K_D = 2 \times I_2, K_I = 1 \times I_2, \Lambda = 2 \times I_2$. Figure 5 represents the results for tracking performance of Lyapunov-based controller with uncertainty observer.

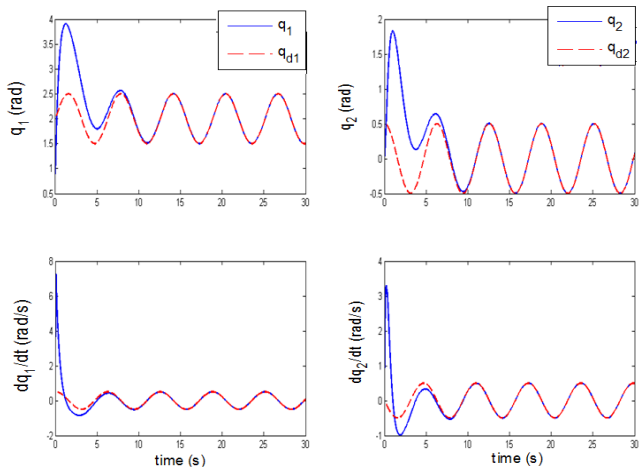

Fig. 5. Trajectory performance based on Lyapunov-based controller with uncertainty observer in the presence of tremor

Figure 6. shows practical result for 5 DOF surgical robot that is designed in this paper. It is mentioned that by using proposed control method, tracking error converges to zero finally.

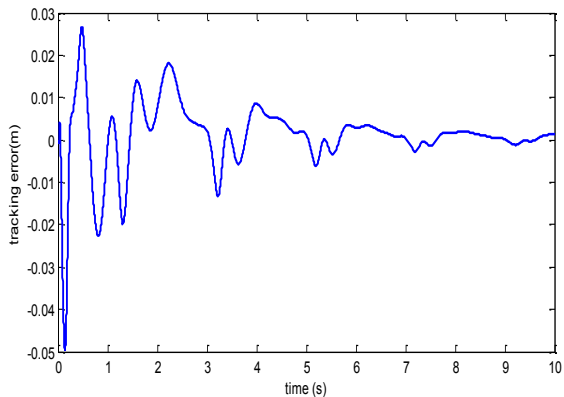

Fig. 6 Tracking error signal for practical robot

As it can be seen, controller performance is robust against variation of uncertainty magnitude. Lyapunov-based controller does not make use of equation decoupling or nonlinear terms elimination, so it is more robust to parameter uncertainties in comparison with inverse-dynamic controllers.

## 6. Conclusion

This paper focuses on mechanical designing, manufacturing, dynamic analyzing and controlling of the surgical robot's arm with 5 degree of freedom. First, the design and manufacturing limitations of one arm is considered and due to these studies, arm is designed and manufactured based on spherical coordinates. Dynamic analysis for a simplified model is done and a controller is designed based on it. It is proposed that due to compactness of the mechanical structure of arm and the need for laparoscopic surgeries, this arm can be installed on a robot with 2 degrees of freedom in Cartesian coordinates.

Lyapunov –based controller with uncertainty observer is presented in this paper to improve the accuracy of tracking procedure for surgical manipulators with constant bounded disturbances.

By comparing inverse dynamics controller with an integral action and Lyapunov –based controller with uncertainty observer, it can be stated that inverse dynamic controller has the capability to be modified, whilst Lyapunov-based controller cannot be easily modified due to its structural complexities. But from another viewpoint, the Lyapunov-based controller is efficiently robust to uncertainties due to its nonlinear nature, where inverse dynamic controller is very sensitive to parameter uncertainties.

As a practical example, a 5 degree of freedom surgical robot is used to assess the performance of proposed methods. Simulation results confirm that control schemes provide an effective means of obtaining high performance trajectory tracking and show good parameter convergence.

## References

[1] R. Muradore, D. Bresolin, L. Geretti, P. Fiorini, T. Villa, "Robotic surgery" *Robotic and Automation Magazine, IEEE*, Volume 18, Issue 3, 2011, Pages24-32.

[2] I. S. Gill, (2006). "Textbook of laparoscopic urology". Informa Healthcare USA, Inc.

[3] R. Abovitz. (2001). "Digital surgery: the future of medicine and human-robot symbiotic interaction". *Industrial Robot: An International Journal*,Vol.28,No.5.pp.401-405.

[4] M. C. Cavusoglu, F.Tendick, M. Cohn, S. S. Sastry. (1999)."A Laparoscopic Telesurgical Workstation", *IEEE Transactions on Robotics and Automation*.

[5] H. Kumon, M. Murai, S. Baba. " Endourooncology: new horizons in endourology". Springer. pp 39-46: 27-38.

[6] M. Hadavand, A. Mirbagheri, H. Salarieh, F. Farahmand "Design of a force-reflective master robot for haptic telesurgery applications:Robomaster1" *Engineering in Medicine and Biology Society,Annual International Conference of the IEEE*, 2011, Pages 305-310.

[7] Kim, K.-Y.; Song, H.-S.; Suh, J.-W.; Lee, J.-J., "A Novel Surgical Manipulator with Workspace-Conversion Ability for Telesurgety" *Mechatronics, IEEE/ASME Transactions*, Volume 18, Issue 1, 2013, Pages 200-211.

[8] M. Tavakoli, R V. Patel, M.Moallem, A.Aziminejad, *Haptics for Teleoperated Surgical Robotics Systems*. World Scientific Publishing Co. Pte. Ltd, 2008.

[9] Ji Ma, P. Berkelman, "Control Software Design of A Compact Laparoscopic Surgical Robot System" *IEEE International Conference on Intelligent Robots and Systems* , 2006, pp 2345-2350.

[10] P. E. Sandin. (2003). "Robot mechanisms and Mechanical Devices".

[11] J. Craig. (2005). "Introduction to Robotics-Mechanics and Control". *The McGraw-Hill publication. Third Edition*.

[12] M. Wilson, "The role of seam tracking in robotic welding and bonding", The Industrial Robot. vol. 29, no. 2, pp.132-137, 2002.

[13] P. E. Wellstead ,M. B. Zarrop, Self-Tuning Systems: Control and Signal Processing, John Wiley a Sons,1991.

[14] A. Aminzadeh Ghavifekr, S. Pezeshki, A. Arjmandi, "Evaluation of Three Nonlinear Control Method to reject Constant Bounded Disturbance for Robotic Manipulators" *Majlesi Journal of Mechatronic, 2012, Vol 1, No2, pp 38-46*.

[15] U. Bauch, J. Bliedtner , H. Muller, M. Pitzschler, G. Staupendahl. (2000). "Precision processing of composites: laser beam and water jet abration techniques". Wiley, Volume 2, Issue 3.

[16] L. Yahia. (2010)."The Effects of Steam Sterilization on Stainless Steel Instruments". LiAB, Laboratoired' Innovation et d'Analyse de Bioperformance, & ECOLE POLYTECHNIQUE M O N T R E A L.

[17] Mark W.Spong, Seth Hutchinson, M Vidyasagar, Robot Modeling and Control. Wiley Press, 2006

[18] H. Khalil, Nonlinear systems, Third edition, Prentice hall, 2001.